\documentclass[11pt]{article}

\usepackage{amsmath, amssymb, amsthm, verbatim,enumerate,bbm,color} 
\numberwithin{equation}{section}
\usepackage{float}

\usepackage{mathrsfs}

\usepackage{tikz}
\usetikzlibrary{calc}
\usetikzlibrary{math}
\usetikzlibrary{decorations,decorations.markings,decorations.pathreplacing,decorations.pathmorphing}
\usetikzlibrary{fadings}
\usetikzlibrary{fit}
\usetikzlibrary{matrix}
\usetikzlibrary{patterns}
\usetikzlibrary{positioning}
\usetikzlibrary{shadows}
\usetikzlibrary{shapes,shapes.geometric}
\usetikzlibrary{trees}
\usetikzlibrary{overlay-beamer-styles}
\usepackage{tikzsymbols}

\usepackage[colorlinks,citecolor=blue,bookmarks=true]{hyperref}

\date{\today}
\allowdisplaybreaks

\theoremstyle{plain}
\newtheorem{theorem}{Theorem}[section]

\newtheorem{conjecture}[theorem]{Conjecture}
\newtheorem{problem}[theorem]{Problem}

\newtheorem{definition}[theorem]{Definition}

\makeatletter
\def\moverlay{\mathpalette\mov@rlay}
\def\mov@rlay#1#2{\leavevmode\vtop{%
		\baselineskip\z@skip \lineskiplimit-\maxdimen
		\ialign{\hfil$\m@th#1##$\hfil\cr#2\crcr}}}
\newcommand{\charfusion}[3][\mathord]{
	#1{\ifx#1\mathop\vphantom{#2}\fi
		\mathpalette\mov@rlay{#2\cr#3}
	}
	\ifx#1\mathop\expandafter\displaylimits\fi}
\makeatother



\RequirePackage{color}\definecolor{RED}{rgb}{1,0,0}\definecolor{BLUE}{rgb}{0,0,1} 

\newcommand{\poly}{\text{poly}}

\usepackage{hyperref} 

\title{Polynomial Property Testing}

\date{}

\author{
	Lior Gishboliner\thanks{University of Toronto, Canada. Email: lior.gishboliner@utoronto.ca. 
	Supported by the NSERC Discovery Grant ``Problems in Extremal and Probabilistic Combinatorics".	
	In the preliminary stages of writing this survey, LG was supported by SNSF grant 200021\_196965.} 
	\and Asaf Shapira\thanks{School of Mathematics, Tel Aviv University, Tel Aviv 69978, Israel. Email: asafico@tau.ac.il. Supported in part by ERC Consolidator Grant 863438 and NSF-BSF Grant 20196.} }

\begin{document}
	
	\maketitle
	
	
	\vspace{1cm}
	
	\begin{abstract}
		Property testers are fast, randomized ``election polling"-type algorithms that determine if an input (e.g., graph or hypergraph) has a certain property or is $\varepsilon$-far from the property. In the dense graph model of property testing, it is known that many properties can be tested with query complexity that depends only on the error parameter $\varepsilon$ (and not on the size of the input), but the current bounds on the query complexity grow extremely quickly as a function of $1/\varepsilon$. Which properties can be tested {\em efficiently}, i.e., with $\poly(1/\varepsilon)$ queries? This survey presents the state of knowledge on this general question, as well as some key open problems.
	\end{abstract}
	
	\section{Introduction}\label{sec:introduction}
	Property testing is an area of theoretical computer science that deals with the design of ultrafast randomized algorithms for determining if an input object has a certain property or is far from having the property. These are approximate decision algorithms which can be thought of as ``election polling". Indeed, the most basic task of this type is, given a binary string of length $n$ and some $\alpha \in (0,1)$, to distinguish between the case that the string has at least $\alpha n$ ones and the case that it has at most $(\alpha-\varepsilon)n$ ones, where $\varepsilon$ is called the {\em error parameter}. In other words, we have to distinguish between the case that a certain party received at least an $\alpha$-fraction of the votes and the case that it received at most an $(\alpha-\varepsilon)$-fraction. Probabilistic concentration inequalities (e.g., Hoeffding's inequality) tell us that with high probability (say, at least $0.99$), a sample of size $O(1/\varepsilon^2)$ will have the same proportion of ones as the entire string up to an error of $\frac{\varepsilon}{2}$, allowing us to estimate the total fraction of ones based on the sample. A crucial point is that the sample size depends only on $\varepsilon$ and not on the size $n$ of the input. 
	
	Property testing applies the above paradigm to ``polling" more involved properties of various combinatorial structures, such as graphs, hypergraphs, binary matrices, permutations, etc. The field emerged from the seminal works of Rubinfeld and Sudan \cite{Rubinfeld_Sudan}, Blum, Luby and Rubinfeld \cite{BLR}, and Goldreich, Goldwasser and Ron \cite{GGR} in the 90s, and has grown significantly in the last three decades. 
	
	\subsection{The dense graph model}
	
	Let us now define precisely what we mean by testing. 
	In this survey we focus on the so-called {\em dense graph model} (also known as the {\em adjacency-matrix model}), where an input graph $G$ is given via its adjacency matrix. A {\em graph property} is simply a family of graphs closed under isomorphism. The distance of a graph $G$ to satisfying a given graph property $\mathcal{P}$ is the fraction of adjacency-matrix entries that need to be changed in order to turn $G$ into a graph satisfying $\mathcal{P}$. 
	This notion of distance is called {\em edit distance}, and
	extends naturally to $k$-uniform hypergraphs, which are given by $k$-dimensional adjacency arrays. Thus, we have the following definition:
	\begin{definition}[$\varepsilon$-close/far]\label{def:far}
		Let $G$ be an $n$-vertex $k$-uniform hypergraph, and let $\varepsilon > 0$. We say that $G$ is {\em $\varepsilon$-close} to a hypergraph property $\mathcal{P}$ if one can turn $G$ into a hypergraph satisfying $\mathcal{P}$ by adding/deleting at most $\varepsilon n^k$ edges. Otherwise, $G$ is {\em $\varepsilon$-far} from $\mathcal{P}$. 
	\end{definition} 
	
	We stress that the definitions presented in this section hold generally for $k$-uniform hypergraphs. Still, in order to keep the presentation simple, we opted to sometimes state definitions only for the case of graphs. In these cases, the generalization to hypergraphs should be evident. 
	
	Note that the adjacency-matrix model is suitable for studying dense graphs, i.e., graphs with $\Theta(n^2)$ edges (or, more generally, $k$-uniform hypergraphs with $\Theta(n^k)$ edges). Indeed, graphs with $o(n^2)$ edges are indistinguishable from the empty graph according to the above definition (in the sense that they are $o(1)$-close to it).
	At this point we would like to point out that there are also other well-studied property testing models, such as the {\em bounded-degree model} (for bounded degree graphs) and the {\em general graph model}. We refer the reader to the excellent book of Goldreich \cite{Goldreich_book} for an overview of these models as well as a general introduction to property testing.

	Returning to the dense graph model, a {\em tester} for a property $\mathcal{P}$ is a randomized algorithm that distinguishes (with high probability) between inputs satisfying $\mathcal{P}$ and inputs which are far from $\mathcal{P}$. The tester can make vertex samples and query the input's adjacency matrix on pairs of sampled vertices. The {\em sample-complexity} of a tester is the number of sampled vertices, and the {\em query-complexity} (which is at most quadratic in the sample-complexity) is the number of edge-queries made. By a result of Goldreich and Trevisan \cite{GT_canonical}, if a property has a tester making $q$ queries, then it is also testable by a tester that works by sampling $O(q)$ vertices and making its decision based on the sample; such a tester is called canonical. Thus, from now on, we will mostly restrict ourselves to canonical testers (at the price of at most squaring the query-complexity). 
	Note that such testers are in particular {\em non-adaptive}, i.e., they make all of their queries at once (instead of basing later queries on the results of previous queries). 
	See the papers of Blais and Seth \cite{BS2} and Goldreich and Ron \cite{GoldreichRon} for some results on cases where there exist testers with better query complexity than the canonical \nolinebreak one.
	
	\begin{definition}[testable]\label{def:testable}
		A property $\mathcal{P}$ of $k$-uniform hypergraphs is {\em testable} if there is a function $q_{\mathcal{P}} : (0,1) \rightarrow \mathbb{N}$ and a canonical tester that, given an error parameter $\varepsilon > 0$ and an input $k$-uniform hypergraph $G$, samples a set of $q_{\mathcal{P}}(\varepsilon)$ vertices from $G$ uniformly at random, and:
		\begin{enumerate}
			\item accepts $G$ with probability at least $\frac{2}{3}$ if $G$ satisfies $\mathcal{P}$;
			\item rejects $G$ with probability at least $\frac{2}{3}$ if $G$ is $\varepsilon$-far from $\mathcal{P}$. 
		\end{enumerate}
		A tester has {\em one-sided error} if it accepts with probability 1 every input that satisfies $\mathcal{P}$. Otherwise the tester has {\em two-sided error}.
	\end{definition}
	\noindent
	A crucial point in Definition \ref{def:testable} is that the sample-complexity $q_{\mathcal{P}}(\varepsilon)$ depends only on $\varepsilon$ (and $\mathcal{P}$) but not on the size of the input. We note that the success probability, which is usually (somewhat arbitrarily) chosen to be $\frac{2}{3}$, can be amplified to $1-\alpha$ by repeating the experiment $\Theta(\log \frac{1}{\alpha})$ times and deciding by majority. 
	
	As we shall see, many natural graph (and hypergraph) properties are testable, but the current upper bounds for the sample-complexity of these testers grow extremely fast with $1/\varepsilon$. This is due to using the powerful (but ``costly") {\em Szemer\'edi's regularity lemma} and its generalizations. This situation leads to the question of which properties have testers with {\em polynomial} sample complexity, i.e., which properties can be tested {\em efficiently}. This is the main topic of the current survey.  
	\begin{problem}\label{prob:main}
		Which properties $\mathcal{P}$ can be tested with sample-complexity $q_{\mathcal{P}}(\varepsilon) = \poly(1/\varepsilon)$?
	\end{problem} 
	For a property $\mathcal{P}$ satisfying $q_{\mathcal{P}}(\varepsilon) = \poly(1/\varepsilon)$, we will say that $\mathcal{P}$ is {\em polynomially testable}.
	Before presenting the current state of knowledge on Problem \ref{prob:main}, we first discuss the aforementioned general testability results, namely, the testability of hereditary graph properties. 

	\subsection{Testing hereditary properties: the removal lemma}
	Let us introduce some basic terminology. 
	For a graph family $\mathcal{F}$, a graph $G$ is {\em $\mathcal{F}$-free} if it has no copy of any graph in $\mathcal{F}$, and is {\em induced $\mathcal{F}$-free} if it has no induced (i.e., isomorphic) copy of any graph in $\mathcal{F}$. When $\mathcal{F} = \{F\}$, we say $F$-free and induced $F$-free, respectively.  
	
	A graph property $\mathcal{P}$ is {\em hereditary} if it is closed under removing vertices. Equivalently, $\mathcal{P}$ is hereditary if it can be characterized by forbidden induced subgraphs, i.e., if there is a (possibly infinite) family of graphs $\mathcal{F}$ such that a graph satisfies $\mathcal{P}$ if and only if it is induced $F$-free for every $F \in \mathcal{F}$. Indeed, one simply takes $\mathcal{F}$ to be the set of all graphs not satisfying $\mathcal{P}$. 
	
	Building on a long line of work \cite{RS,AFKS,AS_monotone}, Alon and the second author \cite{AS_hereditary} proved that every hereditary graph property is testable with one-sided error. This result was later extended to $k$-uniform hypergraphs by R\"odl and Schacht \cite{RodlSchacht_hereditary,RodlSchacht_removal}. These results are essentially combinatorial, not algorithmic; the tester is extremely simple: it samples a set of $q_{\mathcal{P}}(\varepsilon)$ vertices and accepts the input if and only if the subgraph induced by the sample satisfies $\mathcal{P}$. 
	As $\mathcal{P}$ is hereditary, it is clear that inputs satisfying $\mathcal{P}$ are accepted with probability $1$. The other direction in the proof of correctness follows from the following deep combinatorial theorem:
	\begin{theorem}[Infinite hypergraph removal lemma]\label{thm:infinite_removal}
			Let $k \geq 2$ and let $\mathcal{F}$ be a (possibly infinite) family of $k$-uniform hypergraphs. For every $\varepsilon > 0$ there are $\delta = \delta_{\mathcal{F}}(\varepsilon) > 0$ and $m = m_{\mathcal{F}}(\varepsilon)$ such that if $G$ is an $n$-vertex $k$-uniform hypergraph which is $\varepsilon$-far from being induced $\mathcal{F}$-free, then there is $F \in \mathcal{F}$ with $|F| \leq m$ such that $G$ contains at least $\delta n^{|V(F)|}$ copies of $F$. 
	\end{theorem} 
	\noindent
	An equivalent reformulation of Theorem \ref{thm:infinite_removal} is as follows:
	\begin{theorem}[Infinite hypergraph removal lemma, sampling form]\label{thm:infinite_removal sampling}
		For every hereditary property $\mathcal{P}$ of $k$-uniform hypergraphs and for every $\varepsilon > 0$, there is $q = q_{\mathcal{P}}(\varepsilon)$ such that if a $k$-uniform hypergraph $G$ is $\varepsilon$-far from $\mathcal{P}$, then with probability at least $0.99$, the subgraph induced by a sample of $q_{\mathcal{P}}(\varepsilon)$ vertices of $G$ does not satisfy $\mathcal{P}$. 
	\end{theorem}
	\noindent
	Theorem \ref{thm:infinite_removal sampling} exactly corresponds to the correctness of the aforementioned tester for $\mathcal{P}$. If $q_{\mathcal{P}}(\varepsilon) = \poly(1/\varepsilon)$ then we will say that $\mathcal{P}$ has a {\em polynomial removal lemma} (or is polynomially testable). 
	
	Theorem \ref{thm:infinite_removal} has a long history. It was first proved in the special case $k=2$ and where the corresponding hereditary property is (not necessarily induced) $F$-freeness for a fixed graph $F$. This is the seminal {\em graph removal lemma} of Ruzsa and Szemer\'edi \cite{RS}. Its proof is one of the first applications of the Szemer\'edi regularity lemma \cite{Szemeredi}. Ruzsa and Szemer\'edi \cite{RS} used this result to prove their famous $(6,3)$-theorem. They also observed that this theorem is surprisingly related to additive combinatorics; it implies Roth's theorem \cite{Roth}, stating that a subset of $[n]$ containing no 3-term arithmetic progression\footnote{This was later extended to progressions of any length in the celebrated theorem of Szemer\'edi \cite{Szemeredi_AP}.} (3-AP for short) has size $o(n)$. This connection also allowed Ruzsa and Szemer\'edi to (implicitly) prove that the triangle-removal lemma (i.e., Theorem \ref{thm:infinite_removal sampling} in the case $k=2$ and $\mathcal{P} = $ triangle-freeness) cannot be tested with polynomial sample complexity. More precisely, they showed that 
	$q_{\mathcal{P}}(\varepsilon) \geq (1/\varepsilon)^{\Omega(\log(1/\varepsilon))}$. This was proved using the connection to sets avoiding 3-APs and the construction of Behrend \cite{Behrend} giving a 3-AP-free subset of $[n]$ of size 
	$n e^{-O(\sqrt{\log n})}$. 
	
	The next major step towards Theorem \ref{thm:infinite_removal} was the {\em induced-removal lemma} of Alon, Fischer, Krivelevich and Szegedy \cite{AFKS}, which corresponds to the case of Theorem \ref{thm:infinite_removal} where $k=2$ and $\mathcal{F}$ is a finite family. Later, Alon and the second author \cite{AS_monotone,AS_hereditary} found a way of proving the theorem also for infinite families $\mathcal{F}$ (for $k=2$). 
	
	The generalization to $k$-uniform hypergraphs ($k \geq 3$) is the result of the development of the {\em hypergraph regularity method}, i.e., of extending Szemer\'edi's lemma to hypergraphs. This major achievement was obtained independently by Gowers \cite{Gowers_hypergraph} and Nagle-R\"odl-Skokan-Schacht \cite{NRS,Rodl_Skokan_1,Rodl_Skokan_2}. Another proof was later given by Tao \cite{Tao}. 
	 
	As mentioned above, the proof of Theorem \ref{thm:infinite_removal} relies on Szemeredi's regularity lemma \cite{Szemeredi} and its generalizations. Roughly speaking, the regularity lemma states that every graph can be partitioned into a bounded number of parts, such that most pairs of parts behave in a ``random-like" fashion. More precisely, a pair of disjoint vertex-sets $X,Y$ in a graph is called {\em $\varepsilon$-regular} if for every $X' \subseteq X, Y' \subseteq Y$ with $|X'| \geq \varepsilon|X|, |Y'| \geq \varepsilon |Y|$, it holds that
	$|d(X',Y') - d(X,Y)| \leq \varepsilon$, where $d(X,Y) := \frac{e(X,Y)}{|X||Y|}$ is the {\em density} of $(X,Y)$. An {\em equipartition} of a set is a partition in which the sizes of any two parts differ by at most $1$. 
	\begin{theorem}[Szemer\'edi regularity lemma]\label{thm:Szemeredi}
		For every $\varepsilon > 0$ there is $T = T(\varepsilon)$ such that every graph $G$ admits an equipartition $V(G) = V_1 \cup \dots \cup V_t$ with $t \leq T$ such that all but at most $\varepsilon t^2$ of the pairs $(V_i,V_j)$ are $\varepsilon$-regular.
	\end{theorem}
	
	The regularity lemma is usually used together with a {\em counting lemma}, which allows one to count copies of a fixed graph $F$ in an appropriate configuration consisting of dense regular pairs. For example, if $V_1,V_2,V_3$ are three vertex-disjoint sets such that all pairs $(V_i,V_j)$ are $\varepsilon$-regular and have density at least $2\varepsilon$, then there are at least $\poly(\varepsilon)|V_1||V_2||V_3|$ triangles with one vertex in each $V_i$. Thus, to prove Theorem \ref{thm:infinite_removal} in the special case $k=2$ and $\mathcal{P} = $ triangle freeness, one takes a regular partition (given by Theorem \ref{thm:Szemeredi}) and ``cleans" it, deleting edges between pairs that are not regular or not dense. Since the cleaning deletes only few edges, the remaining graph still has a triangle, which implies (using the counting lemma) that the graph in fact has at least $\delta n^3$ triangles (for an appropriate $\delta = \delta(\varepsilon)$).

	The proof of the regularity lemma gives an upper bound on $T(\varepsilon)$ of the form $T(\varepsilon) \leq \text{tow}(\poly(1/\varepsilon))$, where $\text{tow}(x)$ is the tower function defined by $\text{tow}(x) = 2^{\text{tow}(x-1)}$. 
	This results in the bound $q_{\mathcal{P}}(\varepsilon) \leq \text{tow}(\poly(1/\varepsilon))$ in Theorem \ref{thm:infinite_removal sampling} even for simple properties\footnote{For properties defined in terms of infinitely many forbidden subgraphs, the bounds in Theorem \ref{thm:infinite_removal sampling} can be arbitrarily large, see \cite{AS_monotone}. See also \cite{GS_cycles} for a treatment of the special case of properties defined by infinitely many forbidden cycles, where tight bounds on $q_{\mathcal{P}}(\varepsilon)$ are proved as a function of the growth rate of the sequence of cycle lengths.}, such as $F$-freeness for a fixed graph $F$. 
	Gowers \cite{Gowers_reg_lemma} showed that such bounds in Theorem \ref{thm:Szemeredi} are unavoidable (see \cite{FoxLovasz} for more precise bounds), meaning that we cannot get improved bounds in Theorem \ref{thm:infinite_removal sampling} by using the Szemer\'edi regularity lemma. 
	An improved bound in Theorem \ref{thm:infinite_removal sampling} was achieved by Fox \cite{Fox} via a proof that avoids the use of the regularity lemma, and by Moshkovitz and the second author \cite{MS_SRAL} by using a weaker form of the regularity lemma with better bounds. Both of these works proved the bound $q_{\mathcal{P}}(\varepsilon) \leq \text{tow}(O(\log 1/\varepsilon))$ for $\mathcal{P} = $ $F$-freeness. It is a major open problem to prove a non-tower-type bound (or show that this is impossible). 
	 
	\paragraph{Paper organization.}The rest of this survey is organized as follows. In Section \ref{sec:partition} we consider partition properties, a large class of properties that includes, e.g., $k$-colorability or properties defined via the maximum cut or maximum clique size. These properties were among the first to be studied in the dense graph model. In Section \ref{sec:hereditary} we consider subgraph-freeness properties and survey the current state of knowledge on removal lemmas with polynomial bounds in graphs and hypergraphs. In Section \ref{sec:ordered} we discuss analogous results for directed and ordered graphs and related structures. Section \ref{sec:estimation} deals with the problem of distance estimation. Finally, in Section \ref{sec:permutations} we discuss the property testing of permutations.

	\section{Partition Properties}\label{sec:partition}
	A {\em partition property} is a property expressing that a graph has a partition (into a given number of parts $k$) with a certain number of edges (and/or a certain density) between parts and within parts, and also with bounds on the sizes of the parts. For example, $k$-colorability is the property of having a partition into $k$ parts which are independent sets. The property of containing a clique of size at least $\alpha n$ can be described as having a partition $V(G) = V_1 \cup V_2$ with $|V_1| \geq \alpha n$ and $d(V_1) = 1$ (where $d(X)$ is the density of $X$, i.e., $d(X) = e(X)/\binom{|X|}{2}$). Another example is having a cut with at least $\alpha n^2$ edges. 
	
	Partition properties were the first properties shown to be testable (and, in fact, polynomially testable) in the seminal paper of Goldreich, Goldwasser and Ron \cite{GGR}. The original definition of partition properties in \cite{GGR} only allowed {\em absolute} bounds on the number of edges, i.e., bounding the number of edges between parts $V_i,V_j$ by a fraction of $n^2$ (and not as a fraction of $|V_i||V_j|$). Later on, Nakar and Ron \cite{NR} considered a more general class of partition properties, allowing also {\em relative} bounds, i.e., bounds on the density between or within sets. This gives the following definition:
	
	\begin{definition}[Partition property]\label{def:partition property}
		A {\em partition property} is given by an integer $k \geq 1$ and real numbers in $[0,1]$ as follows:
		\begin{itemize}
			\item $\rho_i^L \leq \rho_i^U$ for $1 \leq i \leq k$;
			\item $\alpha_{i,j}^L \leq \alpha_{i,j}^U$ for $1 \leq i < j \leq k$ and 
			$\alpha_i^L \leq \alpha_i^U$ for $1 \leq i \leq k$;
			\item $d_{i,j}^L \leq d_{i,j}^U$ for $1 \leq i < j \leq k$ and 
			$d_i^L \leq d_i^U$ for $1 \leq i \leq k$.
		\end{itemize} 
		An $n$-vertex graph $G$ satisfies the property if it has a vertex-partition $V(G) = V_1 \cup \dots \cup V_k$ such that:
		\begin{itemize}
			\item $\rho_i^L n \leq |V_i| \leq \rho_i^U n$ for every $1 \leq i \leq k$;
			\item $\alpha_{i,j}^L n^2 \leq e(V_i,V_j) \leq \alpha_{i,j}^U n^2$ for $1 \leq i < j \leq k$ and $\alpha_i^L n^2 \leq e(V_i) \leq \alpha_i^U n^2$ for $1 \leq i \leq k$. 
			\item $d_{i,j}^L \leq d(V_i,V_j) \leq d_{i,j}^U$ for $1 \leq i < j \leq k$ and $d_i^L \leq d(V_i) \leq d_i^U$ for $1 \leq i \leq k$. 
		\end{itemize}
	\end{definition}
	Namely, the first item above gives bounds on the sizes of the parts, the second item gives (absolute) bounds on the numbers of edges between and within parts, and the last item gives bounds on the densities between and within parts. 
	
	As mentioned above, Goldreich, Goldwasser and Ron \cite{GGR} considered partition properties without density constraints (i.e., without the third Item in Definiton \ref{def:partition property}). They showed that each such partition property is testable with sample complexity $(k/\varepsilon)^{O(k)}$. Nakar and Ron \cite{NR} later found a way of using this result to design a tester for every partition property (allowing density constraints), by reducing density constraints to absolute constraints. Recently, the second author and Stagni \cite{SS_partition} showed that the partition properties considered in \cite{GGR} (without density constraints) can in fact be tested with sample complexity $\poly(k/\varepsilon)$, improving the dependence on $k$ from exponential to polynomial. Combining this with the aforementioned reduction of \cite{NR} gives the following:
	\begin{theorem}[\cite{NR,SS_partition}]\label{thm:partition}
		Every partition property is testable with sample complexity $\poly(k/\varepsilon)$. 
	\end{theorem}
	

%

	Nakar and Ron \cite{NR} also showed that the hereditary partition properties are polynomially testable with {\em one-sided error} (and that no other partition property is testable with one-sided error). The hereditary partition properties are precisely those that do not allow constraints on the sizes of $V_i$ or on the number of edges between parts, and only allow density constraints where $d_{i,j}^L = d_{i,j}^U = 0$ or $d_{i,j}^L = d_{i,j}^U = 1$ or $d_{i,j}^L = 0, d_{i,j}^U = 1$ (and the same for $d_i^L,d_i^U$). In other words, a hereditary partition property is given by a function $d : [k]^2 \rightarrow \{0,1,\perp\}$, and a graph satisfies the property if it has a partition $V_1 \cup \dots \cup V_k$ such that $(V_i,V_j)$ is complete if $d(i,j) = 1$ and empty if $d(i,j) = 0$, where $i=j$ is also allowed; if $d(i,j) = \perp$ then there are no constraints on $(V_i,V_j)$. 
	
	Testing hereditary partition properties is a special case of the much more general framework of testing {\em satisfiability of constraint-satisfaction problems (CSPs)}, which we discuss in the next section. 
	
	\subsection{Testing satisfiability}
	
	Throughout this section, we fix two integer parameters $r,k \geq 2$, where $r$ is called the {\em arity} and $k$ the {\em alphabet size}. 
	
	\begin{definition}[CSP, satisfiable]
		A {\em constraint} with variables $x_1,\dots,x_r$ is a function $f : [k]^r \rightarrow \nolinebreak \{0,1\}$. The constraint is {\em satisfied} by assignment $(a_1,\dots,a_r) \in [k]^r$, i.e., by assigning value $a_i$ to variable $x_i$, if $f(a_1,\dots,a_r) = 1$. An {\em $r$-ary constraint satisfaction problem (CSP)} on variables $x_1,\dots,x_n$ is a collection $\Phi$ of $(r+1)$-tuples $(x_{i_1},\dots,x_{i_r},f)$, where $1 \leq i_1 < \dots < i_r \leq n$ and $f$ is a constraint on $x_{i_1},\dots,x_{i_r}$. A CSP $\Phi$ is {\em satisfiable} if there is an assignment $(a_1,\dots,a_n) \in [k]^n$ which satisfies all constraints. 
	\end{definition}
	
	\noindent
	Now we define the distance of a CSP from satisfiability, in analogy with Definition \ref{def:far}. 
	
	\begin{definition}[$\varepsilon$-close to/far from satisfiable]
	An $r$-ary CSP $\Phi$ with $n$ variables is {\em $\varepsilon$-close} to satisfiability if there is an assignment that satisfies all but at most $\varepsilon n^r$ of the constraints of $\Phi$. Otherwise $\Phi$ is {\em $\varepsilon$-far} from satisfiability. 
	\end{definition}
	
	Testing CSP satisfiability is the problem of distinguishing (with high probability) between satisfiable CSPs and those that are $\varepsilon$-far from satisfiability, using a sample of size depending only on $\varepsilon,k,r$. 
	The tester works by sampling variables and inspecting the set of constraints that only use variables from the sample.
	The fact that CSP satisfiability is polynomially testable with one-sided error was first proved by Alon and the second author \cite{AS_SAT}, who gave a bound\footnote{By including variables in the subscript of a $O$-notation, we mean that the implied multiplicative constant may depend on these variables. Thus, for example, $f(x) \leq O_{k,r}(x)$ means that $f(x) \leq Cx$ for a constant $C$ (possibly) depending on $k,r$.} of $O_{k,r}(\frac{1}{\varepsilon^2})$ on the sample complexity.
	Later, Sohler \cite{Sohler} improved this to $\tilde{O}_{k,r}(\frac{1}{\varepsilon})$, where the hidden constant in the $\tilde{O}$-notation\footnote{We use $\tilde{O}(x)$ to hide factors that are polylogarithmic in $x$.} is roughly $k^r$. Recently, this was improved to a polynomial dependence on both $k$ and $r$ by Blais and Seth \cite{BS2}.
	\begin{theorem}[\cite{BS2}]\label{thm:SAT}
		CSP satisfiability is testable with one-sided error with sample complexity \nolinebreak $\tilde{O}(\frac{kr^3}{\varepsilon})$. 
	\end{theorem}
	Interestingly, the proof in \cite{BS2} uses a new technique (first used by Blais and Seth in \cite{BS1}) of applying the hypergraph container method\footnote{This is a powerful technique with many applications in extremal and probabilistic combinatorics, see \cite{BMS_Containers,ST_Containers}.} to problems in property testing. 
	
	As mentioned above, it is easy to see that hereditary partition properties are a special case of binary CSPs (i.e., $r=2$). Hence, Theorem \ref{thm:SAT} implies 
	that hereditary partition properties are testable with one-sided error with sample complexity $\tilde{O}(\frac{k}{\varepsilon})$, improving earlier bounds in \cite{NR,Sohler}. Fiat and Ron \cite{FR} proved the incomparable bound 
	$\tilde{O}\left(\frac{\log k}{\varepsilon^7} \right)$ 
	on the sample complexity, though their tester has two-sided error. 
	
	\begin{theorem}[\cite{BS2,FR}]\label{cor:0,1 partitions}
		Every hereditary partition property admits a one-sided-error tester with sample complexity $\tilde{O}(\frac{k}{\varepsilon})$ and a two-sided-error tester with sample complexity $\tilde{O}(\frac{\log k}{\varepsilon^7})$.
	\end{theorem} 
	At this point it is also worth mentioning the work of Avigad and Goldreich \cite{AG}, who showed that for a fixed graph $H$, the property of being a blowup of $H$ (which is a hereditary partition property) can be tested with one-sided error with query complexity $\tilde{O}(\frac{1}{\varepsilon})$.
	
	Another well-studied special case of CSP satisfiability is hypergraph $k$-colorability. A hypergraph is {\em $k$-colorable} if there is a partition of its vertices into $k$ independent sets, where an {\em independent set} is a vertex-set containing no edges. 
	The problem of testing hypergraph colorability has a long history, with several works \cite{GGR,AK,AS_SAT,CS_hypergraph_col,Sohler} proving that $k$-colorability of graphs and, more generally, $r$-uniform hypergraphs, is polynomially testable, with a sequence of improving bounds, culminating in the current best bound of \cite{BS1,BS2}, which is a special case of Theorem \nolinebreak \ref{thm:SAT}.  
	
	\begin{theorem}[\cite{BS2}]\label{thm:hyper_col}
		$k$-colorability of $r$-uniform hypergraphs is testable with one-sided error with sample complexity $\tilde{O}(\frac{kr^3}{\varepsilon})$. 
	\end{theorem}
	
	The above theorems suggest the problem of determining the precise dependence of the sample complexity on $k,r,\varepsilon$. Even for some of the simplest cases, such as testing $k$-colorability of graphs, the precise dependence is not known. Namely, it is not known whether the polylogarithmic factor in $\frac{1}{\varepsilon}$ appearing in Theorem \ref{thm:hyper_col} can be eliminated, and whether the dependence on $k$ is polynomial or polylogarithmic. In fact, to the best of our knowledge, the best current lower bound on the sample complexity for testing $k$-colorability is only $\Omega(\frac{1}{\varepsilon})$, where the constant is independent of $k$  (see the paper of Alon and Krivelevich \cite{AK}); i.e., even a bound of the form $\frac{C_k}{\varepsilon}$ for $C_k$ tending to infinity with $k$ is not known. It is also worth mentioning the work of Bogdanov and Trevisan \cite{BT}, who considered the query complexity (instead of sample complexity) of testing 2-colorability, and showed that any non-adaptive tester requires at least $\Omega(\frac{1}{\varepsilon^2})$ queries (this in particular implies the bound $\Omega(\frac{1}{\varepsilon})$ on the sample complexity obtained in \cite{AK}), and that any adaptive tester requires at least $\Omega(\frac{1}{\varepsilon^{3/2}})$ queries.
	
	\begin{problem}
		Determine the optimal sample complexity for testing graph $k$-colorability. 
	\end{problem} 
	\noindent
	In particular, can $k$-colorability be tested with sample complexity $\text{polylog}(k) \cdot\tilde{O}(\frac{1}{\varepsilon})$? 
	
	Alon, de la Vega, Kannan and Karpinski \cite{ADKK} gave a randomized algorithm that not only tests CSP satisfiability, but also estimates the distance of the input CSP $\Phi$ from satisfiability: 
	\begin{theorem}[\cite{ADKK}]\label{thm:ADKK}
		There is a randomized algorithm which, given an $r$-ary CSP $\Phi$ with $n$ variables and given an error parameter $\varepsilon > 0$, samples $\tilde{O}(\frac{1}{\varepsilon^4})$ variables uniformly at random and estimates up to an additive error of $\varepsilon n^r$ with success probability\footnote{As usual, the success probability can be amplified to $1-\alpha$ by repeating the experiment $\Theta(\log \frac{1}{\alpha})$ times.} at least $\frac{2}{3}$ the maximum number of satisfiable constraints in $\Phi$ .
	\end{theorem}

	A similar result (with sample complexity $\tilde{O}(\frac{1}{\varepsilon^7})$) was obtained by Andersson and Engebretsen \cite{AE}. For the special case $r=2$, the polylogarithmic factors in Theorem \ref{thm:ADKK} were removed by a result of Rudelson and Vershynin \cite{RV}, giving a sample complexity of $O(\frac{1}{\varepsilon^4})$. Note that a special case of a binary CSP is the maxcut of a graph. Thus, this result implies that one can estimate the maxcut of a graph up to additive error $\varepsilon n^2$ by examining the subgraph induced by a sample of size $O(\frac{1}{\varepsilon^4})$. It is worth mentioning the following related problem, raised by Yufei Zhao. 
	\begin{problem}\label{prob:maxcut}
		Prove or disprove the following: For every $\varepsilon > 0$ there is $M = M(\varepsilon) > 0$ such that the following holds. Let $G$ be an $n$-vertex graph with $e(G) \geq M \cdot n$, and let $S$ be a random subset of $V(G)$ obtained by including each element randomly and independently with probability $\frac{1}{2}$. Then with high probability, $\left| \frac{\text{maxcut}(G[S])}{e(G[S])} - \frac{\text{maxcut}(G)}{e(G)} \right| \leq \varepsilon$. 
	\end{problem}
	Namely, Problem \ref{prob:maxcut} asks if $G$ and a random subgraph of $G$ on half of the vertices have roughly the same {\em maxcut density}, i.e., the maxcut divided by the total number of edges. The assumption that $e(G) \gg n$ is necessary, because the statement does not hold, e.g., if $G$ is a disjoint union of $\frac{n}{3}$ triangles. As observed by Zhao, Theorem \ref{thm:ADKK} implies that the answer to Problem \ref{prob:maxcut} is positive if $e(G) \geq n^{1.8}$, say. Indeed, taking $\varepsilon \approx n^{-1/4}$, we apply Theorem \ref{thm:ADKK} to both $G$ and $G[S]$ (where $S$ is a random subset of $V(G)$ sampled with probability $\frac{1}{2}$), observing that taking $S$ at random and then\footnote{This double-sampling trick was already used by Goldreich, Goldwasser and Ron \cite{GGR}.} sampling a random subset of $S$ (as done by the algorithm when running on $G[S]$) is the same as sampling a random subset of $V(G)$ (as done by the algorithm when running on $G$). 
	Thus, for a typical $S$, when the algorithm runs on $G[S]$ it (typically) correctly approximates $\frac{\text{maxcut}(G)}{n^2}$, as well as correctly approximating 
	$\frac{\text{maxcut}(G[S])}{|S|^2} \approx \frac{\text{maxcut}(G[S])}{(n/2)^2}$.
	This means that a typical $S$ satisfies
	$\left| \frac{\text{maxcut}(G[S])}{(n/2)^2} - \frac{\text{maxcut}(G)}{n^2} \right| \leq 2\varepsilon$. Multiplying this with $\frac{n^2}{e(G)}$ and using that $e(G[S]) \approx \frac{e(G)}{4}$ with high probability and $\varepsilon \ll \frac{e(G)}{n^2}$, gives the conclusion of Problem \ref{prob:maxcut}.

	Finally, we discuss more closely the sample complexity for testing for large cliques, i.e., for testing if a graph has a clique of size at least $\alpha n$ (this is, of course, a special case of partition properties). The optimal sample complexity for this property was obtained by Blais and Seth in \cite{BS1}.
	\begin{theorem}[\cite{BS1}]\label{thm:clique test}
		The property of containing a clique of size at least $\alpha n$ is testable with sample complexity $\tilde{O}(\frac{\alpha^3}{\varepsilon^2})$.
	\end{theorem}
	It is known \cite{FLS} that the bound in Theorem \ref{thm:clique test} is best possible. The proof of Theorem \ref{thm:clique test} also uses the container method. Interestingly, it was shown in \cite{BS2} that the canonical tester given by Theorem \ref{thm:clique test} is not optimal, and there is a (non-canonical) tester making fewer edge-queries. Indeed, the tester of Theorem \ref{thm:clique test} makes $\tilde{O}(\frac{\alpha^6}{\varepsilon^4})$ edge-queries, while \cite{BS2} gives a tester making only $\tilde{O}(\frac{\alpha^5}{\varepsilon^{7/2}})$ edge-queries. 
	
	\section{Testing Subgraph-Freeness}\label{sec:hereditary}

	In this section we review the state of knowledge on Problem \ref{prob:main} for hereditary properties of graphs and hypergraphs. 
	
	\subsection{Not-necessarily-induced subgraphs}
	
	The first work in this direction is the following theorem of Alon \cite{Alon}:
	\begin{theorem}[\cite{Alon}]\label{thm:Alon}
		Let $F$ be a graph and let $\mathcal{P} = F$-freeness. Then $q_{\mathcal{P}}(\varepsilon) = \poly(1/\varepsilon)$ if and only if $F$ is bipartite. 
	\end{theorem}
	
	For bipartite $F$, the fact that $q_{\mathcal{P}}(\varepsilon) = \poly(1/\varepsilon)$ follows from the K\H{o}v\'ari-S\'os-Tur\'an theorem \cite{KST}. Indeed, if a graph $G$ is $\varepsilon$-far from being $F$-free then it trivially has at least $\varepsilon n^2$ edges, which implies (by the K\H{o}v\'ari-S\'os-Tur\'an theorem) that $G$ has at least $\poly(\varepsilon)n^{|V(F)|}$ copies of $F$. 
	
	For non-bipartite $F$, the proof that $q_{\mathcal{P}}(\varepsilon) \gg \poly(1/\varepsilon)$ works by reducing to the case that $F$ is an odd cycle, which admits a construction similar to the Ruzsa-Szemer\'edi \cite{RS} construction for triangles. The proof relies on the notions of {\em homomorphism} and {\em core}, which we now recall.
	\begin{definition}[graph homomorphism]\label{def:hom}
		A {\em homomorphism} from a graph $G$ to a graph $H$ is a map $\varphi : V(G) \rightarrow V(H)$ such that $\varphi(x)\varphi(y) \in E(H)$ for every $xy \in E(G)$. 
	\end{definition}
	
	\begin{definition}[core]\label{def:core}
		The {\em core} of a graph $F$ is the smallest subgraph $K$ of $F$ such that there is a homomorphism from $F$ to $K$.
	\end{definition}
	\noindent
	One can show that the core of $F$ is unique up to isomorphism, see the paper of Hell and Ne\v{s}et\v{r}il \nolinebreak \cite{HN_core}. 
	
	Graph homomorphisms turned out to be an extremely useful basic notion, and are now ubiquitous in extremal graph theory (see, e.g., the book of Lov\'asz \cite{Lovasz} on homomorphisms and graph limits). 
	The notion of a core also turned out to be very useful in various contexts. 
	Hell and Ne\v{s}et\v{r}il \cite{HN_core} recount that this notion was known as early as the 1960s, and was communicated to them by their shared advisor Gert Sabidussi. It was also studied independently under different names by various other authors, see the references in \cite{HN_core}. Finally, we also refer the reader to the book of Hell and Ne\v{s}et\v{r}il \cite{HN_book} for a thorough overview of graph homomorphisms. 
	 
	There are two key properties of a core that are used in the proof of Theorem \ref{thm:Alon}. The first is that if $K$ is the core of $F$, then $q_{F\text{-free}}(\varepsilon)$ and $q_{K\text{-free}}(\varepsilon)$ are polynomially related. This follows from the fact that a graph with $\delta n^{|V(K)|}$ copies of $K$ contains $\poly(\delta)n^{|V(F)|}$ copies of $F$, which in turn follows from the hypergraph generalization of the K\H{o}v\'ari-S\'os-Tur\'an theorem, see \cite{Erdos_KST}. The second key property is that if a graph $G$ is homomorphic to $K$, then every $K$-copy in $G$ must be ``canonical". Namely, suppose that $V(K) = [k]$ and $G$ has a partition $V(G) = V_1 \cup \dots \cup V_k$ such that the map $V_i \mapsto i$ is a homomorphism from $G$ to $K$. Then every $K$-copy in $G$ is of the form $(v_1,\dots,v_k)$ with $v_i \in V_i$ playing the role of $i \in V(K)$. Indeed, if there is a copy of $K$ that misses some set $V_i$, then this corresponds to a homomorphism from $K$ to a proper subgraph of $K$; such a homomorphism does not exist because $K$ is a core. This good ``control" over the form of $K$-copies is crucial for the proof of Theorem \ref{thm:Alon}.

	A $k$-uniform hypergraph $F$ is {\em $k$-partite} if there is a partition $V(F) = V_1 \cup \dots \cup V_k$ such that every edge of $F$ intersects each of the parts.
	As mentioned above, the K\H{o}v\'ari-S\'os-Tur\'an theorem has a natural generalization to hypergraphs, due to Erd\H{o}s \cite{Erdos_KST}. This result implies that if $F$ is a $k$-partite $k$-uniform hypergraph, then $q_{\mathcal{P}}(\varepsilon) = \poly(1/\varepsilon)$ for $\mathcal{P} = F$-freeness. This raises the question of whether the converse is also true, i.e., whether $k$-partiteness is necessary for polynomial testability (in analogy with Theorem \ref{thm:Alon}). This possibility was first raised by Kohayakawa, Nagle and R\"odl \cite{KNR}. Recently, this was proved by the authors.
	\begin{theorem}[\cite{GS_hypergraphs}]\label{thm:hypergraphs}
		Let $F$ be a $k$-uniform hypergraph and let $\mathcal{P} = F$-freeness. Then $q_{\mathcal{P}}(\varepsilon) = \poly(1/\varepsilon)$ if and only if $F$ is $k$-partite. 
	\end{theorem}
	The proof of Theorem \ref{thm:hypergraphs} also relies on the notion of a core and the aforementioned key properties (which generalize naturally to hypergraphs). 
	
	\subsection{Induced subgraphs}
	Alon and the second author \cite{AS_induced} were the first to study the behavior of $q_{\mathcal{P}}(\varepsilon)$ for $\mathcal{P} = $ induced $F$-freeness (for a fixed graph $F$). They gave an almost complete characterization of the graphs $F$ for which $q_{\text{ind-} F\text{-free}}(\varepsilon) = \poly(1/\varepsilon)$, missing only the graphs $P_4$, $C_4$ and $\overline{C_4}$. Here, $P_k$ (resp. $C_k$) denotes the path (resp. cycle) with $k$ vertices, and $\overline{F}$ is the complement of $F$. Note that $q_{\text{ind-}\overline{F}\text{-free}}(\varepsilon) = q_{\text{ind-}F\text{-free}}(\varepsilon)$. Alon and Fox \cite{AF} later settled the case of $P_4$. Thus, we have the following theorem:
	\begin{theorem}[\cite{AS_induced,AF}]
		Let $F$ be a graph.
		\begin{enumerate}
			\item If $F \in \{P_2,\overline{P_2},P_3,\overline{P_3},P_4\}$, then $q_{\text{ind-} F\text{-free}}(\varepsilon) = \poly(1/\varepsilon)$.
			\item If $F$ is none of the graphs in the previous item, and also $F \neq C_4,\overline{C_4}$, then $q_{\text{ind-} F\text{-free}}(\varepsilon) \geq (1/\varepsilon)^{\Omega(\log 1/\varepsilon)}$. 
		\end{enumerate}
	\end{theorem}
	\noindent
	The above theorem leaves the case $F = C_4$. 
	\begin{conjecture}\label{conj:C4}
		$q_{\text{ind-} C_4\text{-free}}(\varepsilon) = \poly(1/\varepsilon)$. 
	\end{conjecture}
	The authors \cite{GS_C4} made progress towards Conjecture \ref{conj:C4} by proving that $q_{\text{ind-} C_4\text{-free}}(\varepsilon) \leq 2^{\poly(1/\varepsilon)}$.  
	One related property to induced $C_4$-freeness is chordality, i.e., not containing any induced cycle of length at least $4$. De Verclos \cite{Verclos} proved that $q_{\mathcal{P}}(\varepsilon) = \poly(1/\varepsilon)$ for $\mathcal{P} = $ chordality. Regarding other well-studied graph properties, Alon and Fox \cite{AF} proved that $q_{\mathcal{P}}(\varepsilon)\geq (1/\varepsilon)^{\Omega(\log 1/\varepsilon)}$ for $\mathcal{P} = $ the set of perfect graphs and $\mathcal{P} = $ the set of comparability graph.

	The behavior of $q_{\mathcal{P}}(\varepsilon)$ for $\mathcal{P} = $ induced $F$-freeness was also studied for hypergraphs. In \cite{AS_hypergraphs}, Alon and the second author proved that for a $k$-uniform $F$ with $k \geq 3$, the function $q_{\text{ind-} F\text{-free}}(\varepsilon)$ is not polynomial unless $|V(F)| = k$ (i.e., $F$ is an edge or a non-edge; this is the trivial case) or $k = 3$ and $F$ is the $3$-uniform hypergraph $D$ with $4$ vertices and $2$ edges. In \cite{GT_hypergraph}, the first author and Tomon completed the picture by showing that $q_{\text{ind-} D\text{-free}}(\varepsilon) = \poly(1/\varepsilon)$. Thus, we have the following characterization:
	\begin{theorem}[\cite{AS_hypergraphs,GT_hypergraph}]
			Let $F$ be a $k$-uniform hypergraph with $k \geq 3$. Then 
			$q_{\text{ind-} F\text{-free}}(\varepsilon) = \poly(1/\varepsilon)$ if and only if $|V(F)| = k$ or $k = 3, |V(F)| = 4, |E(F)| = 2$. 
	\end{theorem}
	
	Next we mention a result on multicolored graphs. A {\em $k$-colored graph} is a coloring of the edges of a complete graph with $k$ colors; thus, a $2$-colored graph is the same as a graph. The definition of testability and Theorem \ref{thm:infinite_removal} extend naturally to $k$-colored graphs (here the distance is measured in terms of the number of edge-color-changes necessary to attain the given property). 
	The first author \cite{G_Gallai} characterized the $k$-colored graphs $F$, $k\geq 3$, for which $q_{F\text{-free}}(\varepsilon) = \poly(1/\varepsilon)$; the only cases with polynomial dependence on $\varepsilon$ are the case $|V(F)|=2$ (the trivial case) and the case where $k=3$ and $F$ is the rainbow triangle\footnote{Colorings avoiding a rainbow triangle are known as {\em Gallai colorings} and have been studied widely in the graph theory literature, see for example \cite{GallaiCol}.}.
	
	We end this section with two general open problems related to removal lemmas for induced subgraphs. Two subgraphs $F_1,F_2$ of a graph $G$ are called {\em pair-disjoint} if $|V(F_1) \cap V(F_2)| \leq 1$.
	\begin{problem}
		Prove or disprove the following:
		for every graph $F$ and every $\varepsilon > 0$, there is $\delta = \delta_F(\varepsilon) = \poly(\varepsilon) > 0$ such that if an $n$-vertex graph $G$ is $\varepsilon$-far from being induced $F$-free, then $G$ contains a collection of at least $\delta n^2$ pair-disjoint induced\footnote{The analogous statement for non-induced copies is trivial; taking a maximal collection of pair-disjoint $F$-copies and deleting all of their edges makes the graph $F$-free, thus one can take $\delta = \varepsilon/e(F)$.} copies of $F$.
	\end{problem}
	If we drop the requirement that $\delta$ depends polynomially on $\varepsilon$ then the statement follows from Theorem \ref{thm:infinite_removal}, because a graph with $\delta n^{|V(F)|}$ induced copies of $F$ contains a collection of $\frac{\delta}{|V(F)|^2}n^2$ pair-disjoint induced copies of $F$. Indeed, every pair of vertices is contained in at most $n^{v(F) - 2}$ copies of $F$. As a maximal collection of pair-disjoint induced copies of $F$ intersects every other induced $F$-copy in at least 2 vertices, the size of such a collection must be at least $\frac{\delta n^{v(F)}}{n^{v(F)-2}\binom{|V(F)|}{2}} \geq \frac{\delta}{|V(F)|^2}n^2$. 
	
	Theorem \ref{thm:infinite_removal} implies\footnote{To the best of our knowledge, this was first observed by Alon and the second author \cite{AS_monotone}.} that for every pair of graphs $F_1,F_2$ and $\varepsilon > 0$, there is $\delta = \delta_{F_1,F_2}(\varepsilon) > 0$ such that if a graph $G$ is $\delta$-close to being induced $F_i$-free for both $i=1,2$, then $G$ is $\varepsilon$-close to being induced $\{F_1,F_2\}$-free. Indeed, if $G$ is $\varepsilon$-far from being induced $\{F_1,F_2\}$-free then there is $i=1,2$ such that $G$ contains at least $\delta n^{|V(F_i)|}$ induced copies of $F_i$ (for an appropriate $\delta$), which implies that $G$ contains $\Omega(\delta)n^{2}$ pair-disjoint induced copies of $F_i$, and is hence $\Omega(\delta)$-far from being induced $F_i$-free. Does the same hold for $\delta$ that is polynomial in $\varepsilon$?
	\begin{problem}
		Prove or disprove the following: for every pair of graphs $F_1,F_2$ and $\varepsilon > 0$, there is $\delta = \delta_{F_1,F_2}(\varepsilon) = \poly(\varepsilon) > 0$, such that if a graph $G$ is $\delta$-close to being induced $F_i$-free for both $i=1,2$, then $G$ is $\varepsilon$-close to being induced $\{F_1,F_2\}$-free.
	\end{problem}
	
	\subsection{Testing graph properties of bounded VC-dimension}
	There are general results proving polynomial testability of hereditary properties, that rely on an ultra-strong regularity lemma due to Lov\'asz-Szegedy \cite{LS} and independently Alon-Fischer-Newman \cite{AFN} (see also the paper of Fox, Pach and Suk \cite{FPS_VC}). To introduce this result, we need some definitions. 
	\begin{definition}[bi-induced copy]
		Let $H$ be a bipartite graph with parts $A,B$. A {\em bi-induced copy} of $H[A,B]$ in a graph $G$ is an injection $\varphi : V(H) \rightarrow V(G)$ such that for every $a \in A, b \in B$, $ab \in E(H)$ if and only if $\varphi(a)\varphi(b) \in E(G)$.
	\end{definition}
	\noindent
	In other words, a bi-induced copy is an induced copy of the bipartite graph $H[A,B]$ (with no requirements on the edges inside $\varphi(A)$ and $\varphi(B)$). A pair of disjoint vertex-sets $X,Y$ in a graph is {\em $\varepsilon$-homogeneous} if $d(X,Y) \leq \varepsilon$ or $d(X,Y) \geq 1-\varepsilon$. 
	
	\begin{theorem}[Ultra-strong regularity lemma \cite{LS,AFN}]\label{thm:AFN}
		For every $k \geq 2$ and $\varepsilon > 0$, there is $T = T(k,\varepsilon) = \poly(1/\varepsilon)$ such that the following holds. Let $H$ be a $k$-vertex bipartite graph with parts $A,B$. For every $n$-vertex graph $G$, either $G$ contains at least $(n/T)^k$ bi-induced copies of $H[A,B]$, or $G$ has an equipartition $V(G) = V_1 \cup \dots \cup V_t$ with $t < T$ such that all but at most $\varepsilon t^2$ of the pairs $(V_i,V_j)$ are $\varepsilon$-homogeneous. 
	\end{theorem}

	Let us compare Theorem \ref{thm:AFN} with Szemer\'edi's regularity lemma (Theorem \ref{thm:Szemeredi}). In Theorem \ref{thm:AFN}, the pairs $(V_i,V_j)$ are $\varepsilon$-homogeneous (which is stronger than $\varepsilon$-regular\footnote{Indeed, it is easy to check that an $\varepsilon$-homogeneous pair is $\varepsilon^{1/3}$-regular, say.}). Also, crucially, the number of parts in Theorem \ref{thm:AFN} depends only polynomially on $\varepsilon$ (with the power of the polynomial depending on $k$), whereas the number of parts in Theorem \ref{thm:Szemeredi} can be of tower-type. Thus, graphs with no (or only few) bi-induced copies of a given bipartite graph $H$ have very efficient regularity \nolinebreak partitions. 
	
	Theorem \ref{thm:AFN} is closely related to the notion of VC-dimension. For a set-system $\mathcal{A}$ on $[n]$, the {\em VC-dimension} of $\mathcal{A}$ is the largest $d$ for which there is a subset $X \subseteq [n]$ of size $d$ such that 
	$\{X \cap A : A \in \mathcal{A}\} = 2^X$, i.e., every subset of $X$ is attained as the intersection of $X$ with some set in $\mathcal{A}$. This notion plays an important role in computer science (especially in learning theory), combinatorics, and discrete geometry. For a graph $G$, one considers the set-system on $V(G)$ consisting of all neighborhoods of vertices in $G$, i.e., $\mathcal{A} = \{N(v) : v \in V(G)\}$. It is easy to see that if this set system has VC-dimension at least $k + \log_2(2k)$, then $G$ contains a bi-induced copy of every $k \times k$ bipartite graph.\footnote{Indeed, if the VC dimension of $G$ is at least $k + \log_2(2k)$, then we can find a set $X$ of size at least $k + \log_2(2k)$ and vertices $v_I \in V(G)$ for $I \subseteq X$ such that $N(v_I) \cap X = I$. Fix any $Y \subseteq X$ of size $k$. For each $J \subseteq Y$, there are at least $2^{|X|-|Y|} \geq 2k$ vertices $v \in V(G)$ with $N(v) \cap Y = J$. Hence, we can pick $k$ distinct vertices outside $Y$ with any desired neighborhoods in $Y$. This gives a bi-induced copy of any $k \times k$ bipartite graph.} 
	This is the connection to Theorem \ref{thm:AFN}.
	In fact, the proof of Theorem \ref{thm:AFN} (implicitly) uses one of the key properties of VC-dimension, namely the lemma of Haussler \cite{Haussler} stating\footnote{This is in turn closely related to the famous Sauer-Shelah lemma, which states that a set-system of bounded VC-dimension contains only polynomially many sets.} that in a set-system on $[n]$ with VC dimension $d$, the maximum number of sets at pairwise (hamming) distance at least $\varepsilon n$ is at \nolinebreak most \nolinebreak $(\frac{C}{\varepsilon})^{d}$.
	
	Alon, Fischer and Newman \cite{AFN} used Theorem \ref{thm:AFN} to prove a polynomial removal lemma for {\em bipartite host graphs}. Instead of adapting Definitions \ref{def:far} and \ref{def:testable} to this setting, we state this result directly.
	For a bipartite graph $H$ with parts $A,B$ and a bipartite graph $G$ with parts $X,Y$, an {\em induced copy of $H[A,B]$ in $G[X,Y]$} is a bijection $\varphi : V(H) \rightarrow V(G)$ with $\varphi(A) \subseteq X, \varphi(B) \subseteq Y$, such that for every $a \in A, b\in B$, $ab \in E(H)$ if and only if $\varphi(a)\varphi(b) \in E(G)$.  
	\begin{theorem}[\cite{AFN}]\label{thm:AFN_bipartite}
		Let $\mathcal{H}$ be a finite family of bipartite graphs. For every $\varepsilon > 0$ there is $\delta = \delta(\varepsilon) = \poly_{\mathcal{H}}(\varepsilon) > 0$ such that the following holds. 
		Let $G[X,Y]$ be an $n \times n$ bipartite graph, and suppose that at least $\varepsilon n^2$ edges between $X,Y$ must be added/deleted to make $G$ induced $H[A,B]$-free for every $H[A,B] \in \mathcal{H}$. Then there is $H[A,B] \in \mathcal{H}$ such that $G[X,Y]$ has at least $\delta n^{|V(H)|}$ induced copies of $H[A,B]$. 
	\end{theorem}  
	\noindent
	Fischer and Rozenberg \cite{Fischer_Rozenberg} showed that Theorem \ref{thm:AFN_bipartite} does not generalize to more than 2 colors, i.e., to $r$-edge-colorings of complete bipartite graphs $X \times Y$ with $r \geq 3$.

	Lov\'asz and Szegedy \cite{LS} (see also \cite{GS_SA}) observed that a graph $G$ has bounded VC-dimension (or, equivalently, avoids bi-induced copies of some fixed bipartite graph) if and only if $G$ avoids induced copies of some bipartite graph, some co-bipartite graph, and some split graph. Here, a {\em co-bipartite graph} is the complement of a bipartite graph, and a {\em split graph} is a graph whose vertex-set can be partitioned into an independent set and a clique. Thus, these three graph classes (bipartite, co-bipartite, split) capture all ways of partitioning a graph into two sets, each of which is independent or a clique. The authors \cite{GS_SA} used this connection to prove a polynomial removal lemma for induced $\mathcal{F}$-freeness for finite graph-families $\mathcal{F}$ containing a bipartite, co-bipartite and split graph. They also proved a necessary condition for polynomial removal, giving the following theorem:
	\begin{theorem}[\cite{GS_SA}]\label{thm:finite families}
		Let $\mathcal{F}$ be a finite family of graphs. 
		\begin{enumerate}
			\item If $\mathcal{F}$ contains a bipartite graph, a co-bipartite graph and a split graph, then 
			$q_{\text{ind-} \mathcal{F}\text{-free}} = \nolinebreak \poly(1/\varepsilon)$. 
			\item If $\mathcal{F}$ contains no bipartite graph or no co-bipartite graph, then $q_{\text{ind-} \mathcal{F}\text{-free}} \geq (1/\varepsilon)^{\Omega(\log 1/\varepsilon)}$. 
		\end{enumerate}
	\end{theorem}
	
	Characterizing the finite families $\mathcal{F}$ for which $q_{\text{ind-} \mathcal{F}\text{-free}} = \poly(1/\varepsilon)$ remains open. Both conditions in Theorem \ref{thm:finite families} (the sufficient condition of Item 1 and the necessary condition of Item 2) cannot be the correct characterization. Indeed, the property of being a split graph is equivalent to being induced $\{C_4,\overline{C_4},C_5\}$-free (see \cite{Golumbic}) and admits a polynomial removal lemma (by Theorem \ref{cor:0,1 partitions}, as being split is a hereditary partition property), but this property clearly does not forbid any split graph. Also, it was shown in \cite{GS_SA} that there is a bipartite graph $F_1$ and a co-bipartite graph $F_2$ such that $\mathcal{F} = \{F_1,F_2\}$ does not admit a polynomial removal lemma. 
	
	\begin{problem}\label{prob:finite families}
		Characterize the finite graph families $\mathcal{F}$ for which 
		$q_{\text{ind-}\mathcal{F}\text{-free}} = \poly(1/\varepsilon)$.
	\end{problem}  

	The first key open case for Problem \ref{prob:finite families} is the case $\mathcal{F} = \{C_4\}$ (i.e., Conjecture \ref{conj:C4}). Indeed, $C_4$ is bipartite and co-bipartite but not split, so $C_4$ does not fall into either item of Theorem \ref{thm:finite families}. 
	
	For infinite families of forbidden induced subgraphs $\mathcal{F}$, it turns out that containing a bipartite, a co-bipartite and a split graph is not sufficient for polynomial testability; counterexamples were given in \cite{GS_SA}. Still, by adding another condition on $\mathcal{F}$, one can recover polynomial testability. 
	To state this result, we need the following definition: for a graph $G$ and a function $f : V(G) \rightarrow \{0,1\}$, an {\em $f$-blowup} of $G$ is any graph obtained from $G$ by replacing each vertex $x \in V(G)$ with a set $V_x$, such that $(V_x,V_y)$ is a complete bipartite graph if $xy \in E(G)$ and an empty bipartite graph if $xy \notin E(G)$, and $V_x$ is a clique if $f(x) = 1$ and an independent set if $f(x) = 0$. 
	\begin{theorem}[\cite{GS_SA}]\label{thm:blowup}
		Let $\mathcal{F}$ be a (possibly infinite) family of graphs. Suppose that:
		\begin{enumerate}
			\item $\mathcal{F}$ contains a bipartite graph, a co-bipartite graph and a split graph.
			\item For every induced $\mathcal{F}$-free graph $G$, there is a function $f : V(G) \rightarrow \{0,1\}$ such that every $f$-blowup of $G$ is also induced $\mathcal{F}$-free.
		\end{enumerate}
		Then $q_{\text{ind-}\mathcal{F}\text{-free}} = \poly(1/\varepsilon)$.
	\end{theorem} 

	Theorem \ref{thm:blowup} was used in \cite{GS_SA} to prove that every {\em semi-algebraic graph property} admits a polynomial removal lemma. Roughly speaking, these are the properties defined by satisfying a system of polynomial inequalities (where vertices are assigned points in euclidean space). We refer the reader to \cite{GS_SA} for the precise definition and the derivation of this result. 
	
	\section{Directed and Ordered Structures}\label{sec:ordered}
	In this section we consider variants of Theorem \ref{thm:infinite_removal} and the problem of polynomial removal for other combinatorial structures, such as digraphs, ordered graphs and matrices. Definitions \ref{def:far} and \ref{def:testable} extend naturally to these structures. A key common feature of the (sometimes conjectured) characterizations of polynomial testability presented in this section is that $F$-freeness is polynomially testable if and only if the core of $F$ is simple in some sense. Here, ``core" is defined in the same way as in Definition \ref{def:core}, with the definition of homomorphism adapted for each of the combinatorial structures considered. This is in analogy to Theorem \ref{thm:Alon}, which can be stated as saying that $q_{F\text{-free}}(\varepsilon) = \poly(1/\varepsilon)$ if and only if the core of $F$ is a single edge. 
	
	\subsection{Digraphs and tournaments}
	Testing of directed graphs\footnote{Digraphs considered here may have anti-directed edges, i.e., pairs $x,y$ where $(x,y),(y,x)$ are both edges, but may not have parallel edges (two edges from $x$ to $y$).} was first studied by Alon and the second author \cite{AS_directed}, who proved a digraph analogue of the Szemer\'edi regularity lemma (Theorem \ref{thm:Szemeredi}) and used this to prove a digraph analogue of the removal lemma (for the property of $D$-freeness for a given digraph $D$). They also characterized the cases where $q_{D\text{-free}}(\varepsilon) = \poly(1/\varepsilon)$. This result relies on the notion of cores for directed graphs, which is a directed analogue of the notion for undirected graphs (Definition \ref{def:core}). A homomorphism from a digraph $G$ to a digraph $H$ is a mapping $\varphi : V(G) \rightarrow V(H)$ which preserves directed edges, i.e., $(\varphi(x),\varphi(y)) \in E(H)$ for every $(x,y) \in E(G)$. An {\em oriented tree} is an orientation of a tree. 
	\begin{theorem}[\cite{AS_directed}]
		Let $D$ be a digraph. Then $q_{D\text{-free}}(\varepsilon) = \poly(1/\varepsilon)$ if and only if the core of $D$ is an oriented tree or a directed cycle of length 2. 
	\end{theorem}
	
	Fox, Yuster and the authors \cite{FGSY} studied the analogous problem for tournaments. A {\em tournament} is an orientation of a complete graph. Thus, when adapting the definition of distance (Definition \ref{def:far}) to tournaments, one does not allow to delete edges, but only to reverse the direction of edges. In other words, the distance of a tournament $G$ to a tournament property $\mathcal{P}$ is the minimal number of edge-reversals needed to turn $G$ into a tournament satisfying $\mathcal{P}$. 
	Again, one can prove a tournament analogue of the removal lemma by using the digraph analogue of the Szemer\'edi regularity lemma. Let $q^{\text{tour}}_{\mathcal{P}}(\varepsilon)$ denote the sample complexity in the tournament-analogue of Theorem \ref{thm:infinite_removal sampling}. A digraph $D$ is called {\em $2$-colorable} if there is a partition $V(D) = A \cup B$ such that $D[A],D[B]$ are acyclic digraphs, where a digraph is {\em acyclic} if it has no directed cycles. 
	In \cite{FGSY}, the following characterization was \nolinebreak proved.
	\begin{theorem}[\cite{FGSY}]
		Let $D$ be an oriented graph. Then $q^{\text{tour}}_{D\text{-free}}(\varepsilon) = \poly(1/\varepsilon)$ if and only if $D$ is $2$-colorable. 
	\end{theorem}
	
	\noindent
	The proof of the ``if"-direction uses the ultra-strong regularity lemma (Theorem \ref{thm:AFN}). 
	
	We end by mentioning a recent work of Kun and Fekete \cite{FK_posets} studying removal lemmas for posets. Here it is convenient to consider posets as transitive digraphs (where an edge $(x,y)$ indicates that $x < y$). It is shown \cite{FK_posets} that posets admit a removal lemma with polynomial bounds, in the sense that for every poset $F$, if an $n$-vertex poset $P$ contains at most $\delta n^{|V(F)|}$ copies of $F$, then one can delete at most $\varepsilon n^2$ edges of $P$ to obtain an $F$-free poset, where $\delta = \poly(\varepsilon)$. 
	
	\subsection{Ordered graphs and matrices}
	An ordered graph is a graph with a linear order on its vertices. The notions of distance and testability (Definitions \ref{def:far} and \ref{def:testable}) extend verbatim to ordered graphs. The only (crucial) difference is that in this setting, subgraphs (and homomorphisms) must respect the vertex order. Thus, a homomorphism from an ordered graph $G$ to an ordered graph $H$ is a graph homomorphism $\varphi : G \rightarrow H$ which is also order-preserving. A copy of an ordered graph $F$ in $G$ is an injective homomorphism from $F$ to $G$. 
	
	Proving an analogue of Theorem \ref{thm:infinite_removal} (or, equivalently, Theorem \ref{thm:infinite_removal sampling}) for ordered graphs turned out to be considerably more difficult than for unordered ones, due to the need of finding a ``regularity scheme" which works well with the vertex order. Such an analogue, the {\em ordered removal lemma}, was finally proved by Alon, Ben-Eliezer and Fischer in \cite{ABF}. As usual, the proof uses a variant of the Szemer\'edi regularity lemma and produces (at least) tower-type bounds. See also the paper of Towsner \cite{Towsner} for a generalization to hypergraphs.  
	
	Here, too, a natural question is for which properties $\mathcal{P}$ of ordered graphs it holds that $q_{\mathcal{P}}(\varepsilon) = \poly(1/\varepsilon)$. The following was conjectured by the first author and Tomon in \cite{GT_ordered}.
	\begin{conjecture}\label{conj:ordered}
		Let $F$ be an ordered graph. Then 
		$q_{F\text{-free}}(\varepsilon) = \poly(1/\varepsilon)$ if and only if the core of $F$ is an ordered forest.
	\end{conjecture} 
	In \cite{GT_ordered} the ``only if" direction of Conjecture \ref{conj:ordered} was proved, and it was also shown that in order to prove the ``if" direction, it suffices
	to prove the case where $F$ itself is an ordered forest. This remains open. The case where $F$ is an ordered matching was proved by the first author and Šimić \cite{ordered_matchings}. 
	
	The first author and Tomon \cite{GT_ordered} studied the analogous problem for induced subgraphs. Note that in addition to symmetry with respect to complementation, ordered graphs also have symmetry with respect to reversing the vertex order. Namely, for the ordered graph $F^{\leftarrow}$ obtained from $F$ by reversing the order, it holds that
	$q_{\text{ind-}F^{\leftarrow}\text{-free}}(\varepsilon) = q_{\text{ind-}F\text{-free}}(\varepsilon)$. The following characterization was obtained in \cite{GT_ordered}. Let $P$ denote the ordered path with vertices $1,2,3$ and edges $13,23$. 
	\begin{theorem}[\cite{GT_ordered}]
		Let $F$ be an ordered graph. Then 
		$q_{\text{ind-}F\text{-free}}(\varepsilon) = \poly(1/\varepsilon)$ if and only if $|V(F)| = 2$ or $F \in \left\{P,P^{\leftarrow},\overline{P},\overline{P^{\leftarrow}}\right\}$. 
	\end{theorem} 

	Binary matrices can be thought of as bipartite graphs with a vertex-order on each of the parts (i.e., the parts correspond to the rows and columns). The aforementioned ordered removal lemma of Alon, Ben-Eliezer and Fischer \cite{ABF} also applies to matrices (giving tower-type bounds). On the other hand, for unordered bipartite graphs, Theorem \ref{thm:AFN_bipartite} gives a removal lemma with polynomial bounds. 
	This leads to the conjecture (first raised in \cite{AFN}) that (ordered) binary matrices also admit a polynomial removal lemma. To avoid ambiguity, we state this conjecture precisely. A copy of a $k \times k$ binary matrix $A$ in a binary matrix $M$ is a sequence of rows $r_1 < \dots < r_k$ and a sequence of columns $c_1 < \dots < c_k$ of $M$ such that $M_{r_i,c_j} = A_{i,j}$ for all $1 \leq i,j \leq k$. 
	\begin{conjecture}\label{conj:matrices}
		For every $k \times k$ binary matrix $A$ and $\varepsilon > 0$, there is $\delta = \delta(k,\varepsilon) = \poly(\varepsilon) > 0$ such that the following holds. Let $M$ be an $n \times n$ binary matrix, and suppose that one has to change at least $\varepsilon n^2$ entries of $M$ to eliminate all copies of $A$ in $M$. Then $M$ contains at least $\delta n^{2k}$ copies of \nolinebreak $A$. 
	\end{conjecture} 
	\noindent
	One can make the same conjecture more generally for finite families of matrices $A$. 
	Conjecture \ref{conj:matrices} is one of the key open problems in the area, and is not even known, e.g., for the $2 \times 2$ identity matrix. 
	Alon and Ben-Eliezer \cite{AB} proved the following weakening of Conjecture \ref{conj:matrices}. Two submatrices of a matrix $M$ are called disjoint if they do not share any entries. 
	
	\begin{theorem}[\cite{AB}]\label{thm:matrices}
		For every $k \times k$ binary matrix $A$ and $\varepsilon > 0$, there is $\delta = \delta(k,\varepsilon) = \poly(\varepsilon) > 0$ such that the following holds. Let $M$ be an $n \times n$ binary matrix, and suppose that $M$ contains a collection of at least $\varepsilon n^2$ pairwise-disjoint copies of $A$. Then $M$ contains at least $\delta n^{2k}$ copies of \nolinebreak $A$.		
	\end{theorem}
	Note that if $M$ has $\varepsilon n^2$ pairwise-disjoint copies of $A$, then clearly one has to change at least $\varepsilon n^2$ entries to destroy all $A$-copies, meaning that the premise of Theorem \ref{thm:matrices} is stronger than that of Conjecture \ref{conj:matrices}. In other words, 
	Theorem \ref{thm:matrices} proves the conclusion of Conjecture \ref{conj:matrices} under a stronger premise. Alon and Ben-Eliezer \cite{AB} also proved that if a finite family $\mathcal{A}$ of matrices is closed under row permutations (but not necessarily under column permutations), then $\mathcal{A}$ satisfies the conclusion of (the finite-family analogue of) Conjecture \ref{conj:matrices}. This extends Theorem \ref{thm:AFN_bipartite}, which can be stated as saying that Conjecture \ref{conj:matrices} holds for finite matrix-families $\mathcal{A}$ closed under both row and column \nolinebreak permutations. 
	
	\section{Testing vs. Estimation}\label{sec:estimation}
	Estimation is the algorithmic task of estimating an input's distance to a given property. In the dense graph model, estimation is defined as follows:
	\begin{definition}[estimable]\label{def:estimation}
		A graph property $\mathcal{P}$ is {\em estimable} if there is a function $r_{\mathcal{P}} : (0,1) \rightarrow \mathbb{N}$ and a canonical tester that, given a distance parameter $\alpha \geq 0$, an error parameter $\varepsilon > 0$, and an input graph $G$, samples $r_{\mathcal{P}}(\varepsilon)$ vertices from $G$ uniformly at random, and:
		\begin{enumerate}
			\item accepts $G$ with probability at least $\frac{2}{3}$ if $G$ is $\alpha$-close to $\mathcal{P}$. 
			\item rejects $G$ with probability at least $\frac{2}{3}$ if $G$ is $(\alpha + \varepsilon)$-far from $\mathcal{P}$. 
		\end{enumerate} 
	\end{definition}
	Note that the case $\alpha = 0$ in Definition \ref{def:estimation} corresponds to testability (i.e., Definition \ref{def:testable}), hence estimation is at least as hard as testability with two-sided error. 
	We note that estimation is also called {\em tolerant testing}.
	In one of the seminal results in the area, Fischer and Newman \cite{FN} proved that testability and estimability are (qualitatively) equivalent:
	\begin{theorem}[\cite{FN}]\label{thm:FN}
		Every testable graph property is estimable. 
	\end{theorem} 
	The proof of Theorem \ref{thm:FN} uses the Szemer\'edi regularity lemma to transform a tester for $\mathcal{P}$ into an estimator. The reliance on the regularity lemma leads to a tower-type increase of the sample complexity. Namely, the proof shows that if $\mathcal{P}$ can be tested with sample complexity $q_{\mathcal{P}}(\varepsilon)$, then $\mathcal{P}$ can be estimated with sample complexity that is (at least) a tower of height $q_{\mathcal{P}}(\varepsilon)$. For hereditary graph properties, this was dramatically improved by Hoppen, Kohayakawa, Lang, Leffman and Stagni \cite{HKLLS_monotone,HKLLS_hereditary} to a bound that is double-exponential in $q_{\mathcal{P}}(\varepsilon)$. More precisely, they showed that for a (possibly infinite) graph-family $\mathcal{F}$, induced $\mathcal{F}$-freeness can be estimated with sample complexity $e^{(1/\delta)^{O(m^2)}}$, where $\delta = \delta_{\mathcal{F}}(\varepsilon)$ and $m = m_{\mathcal{F}}(\varepsilon)$ are given by Theorem \ref{thm:infinite_removal} (applied in the case $k=2$). This was further improved to $2^{\poly(m/\delta)}$ in a recent work of Kushnir and the authors \cite{GKS}, who also proved a doubly exponential bound for general (not necessarily hereditary) graph properties. 
	\begin{theorem}[\cite{GKS}]\label{thm:estimation} 
		{\color{white} linebreak} 
		\begin{enumerate}
			\item For every graph-family $\mathcal{F}$, induced $\mathcal{F}$-freeness is estimable with sample complexity $2^{\poly(m/\delta)}$, where $\delta = \delta_{\mathcal{F}}(\varepsilon/2)$ and $m = m_{\mathcal{F}}(\varepsilon/2)$ are given by Theorem \ref{thm:infinite_removal}.
			\item If a graph property $\mathcal{P}$ is testable with sample complexity $q_{\mathcal{P}}(\varepsilon)$, then $\mathcal{P}$ is estimable with sample complexity 
			$\exp\Big(\poly(\frac{1}{\varepsilon}) \cdot \exp(q_{\mathcal{P}}(\frac{\varepsilon}{2}))\Big)$. 
		\end{enumerate}
	\end{theorem}
	
	The proof of Theorem \ref{thm:estimation} proceeds by adapting the original proof of Fischer and Newman \cite{FN} (i.e., Theorem \ref{thm:FN}) to work with the {\em Frieze-Kannan regularity lemma} instead of Szemer\'edi's regularity lemma. The Frieze-Kannan regularity lemma \cite{FriezeKannan1,FriezeKannan2} provides a partition that is regular in a weaker sense, but where the number of parts depends only exponentially on $\varepsilon$ (and not in a tower-type way). 
	Frieze-Kannan regularity was also used in the previous works \cite{HKLLS_monotone,HKLLS_hereditary}. The proof of Theorem \ref{thm:estimation} (which improves on the bounds in \cite{HKLLS_monotone,HKLLS_hereditary}) uses several additional ideas, as well as Theorem \ref{thm:partition}. 

	The key open problem in the area is the following:
	\begin{problem}
		Prove or disprove the following: If a graph property $\mathcal{P}$ is testable with sample complexity $q_\mathcal{P}(\varepsilon)$, then $\mathcal{P}$ is estimable with sample complexity $q_{\mathcal{P}}(\poly(\varepsilon))$. 
	\end{problem} 
	
	Fiat and Ron \cite{FR} proved that certain properties that are polynomially testable, such as induced $P_{\ell}$-freeness for $\ell=3,4$ and chordality, are also polynomially estimable. They also showed that hereditary partition properties can be estimated with sample complexity $\poly(\frac{\log k}{\varepsilon})$ (cf. Theorem \nolinebreak \ref{cor:0,1 partitions}).
	
	\section{Permutations}\label{sec:permutations}
	This section is concerned with testing properties of permutations. Permutations are somewhat different from relational structures (such as graphs and hypergraphs), so this topic has a different flavor as compared to previous sections. 
	To consider property testers, we need to decide on a notion of distance (metric) between permutations; there are several well-studied metrics. 
	The most natural analogue of the graph edit distance as given by Definition \ref{def:far} is {\em Kendall's tau distance}, \nolinebreak defined \nolinebreak as \nolinebreak follows:
	\begin{definition}[Kendall's tau distance]\label{def:Kendall}
		Let $\sigma, \pi$ be permutations on $[n]$. 
		For a pair $1 \leq i < j \leq n$, we say that $\sigma,\pi$ disagree on $(i,j)$ if $\sigma(i) < \sigma(j)$ and $\pi(i) > \pi(j)$, or $\sigma(i) > \sigma(j)$ and $\pi(i) < \pi(j)$.
		The {\em Kendall tau distance} $d_{\text{KT}}(\sigma,\pi)$ is defined as
		$$
		d_{\text{KT}}(\sigma,\pi) = \frac{1}{\binom{n}{2}} \cdot \# \{ 1 \leq i < j \leq n : \sigma,\pi \text{ disagree on } (i,j)\}.
		$$
	\end{definition} 
	It is well-known that the number of pairs $(i,j)$ on which $\sigma,\pi$ disagree is precisely the number of adjacent transpositions (i.e., switching the values of $\sigma(i)$ and $\sigma(i+1)$ for some $1 \leq i \leq n-1$) needed to turn $\sigma$ into $\pi$. Thus, $d_{KT}$ is, in a way, analogous to the graph edit distance. 
	
	A {\em permutation property} is simply a set of permutations. Here we focus on hereditary properties of permutations. To define this, we first need to define the notion of a subpermutation.
	\begin{definition}[subpermutation]
		A permutation $\pi$ on $[m]$ is a {\em subpermutation} of a permutation $\sigma$ on $[n]$ if there are $1 \leq i_1 < \dots < i_m \leq n$ such that for every 
		$1 \leq j < k \leq m$, $\sigma(i_j) < \sigma(i_k)$ if and only if $\pi(j) < \pi(k)$.
	\end{definition}
	In other words, $\pi$ is a subpermutation of $\sigma$ if the restriction $\sigma|_I$ of $\sigma$ to some subset $I \subseteq [n]$ has the same {\em order pattern} as $\pi$. For a subset $I = \{i_1,\dots,i_m\} \subseteq [n]$, we denote by $\sigma[I]$ the permutation on $[m]$ with the same order pattern as $\sigma|_I$; we call $\sigma[I]$ the {\em subpermutation induced by $I$}. 
	
	As before, a tester for a permutation property $\mathcal{P}$ is a randomized algorithm that distinguishes (with high probability) between permutations satisfying $\mathcal{P}$ and permutations that are $\varepsilon$-far from $\mathcal{P}$. 
	Here, we say that $\sigma$ is {\em $\varepsilon$-far} from $\mathcal{P}$ if the distance between $\sigma$ and $\pi$ is at least $\varepsilon$ for every $\pi \in \mathcal{P}$. This of course depends on our choice of metric, which is for now the Kendall tau distance. 
	
	We focus on testers that work by {\em examining a random subpermutation}. Namely, given an input permutation $\sigma$ on $[n]$, the tester samples a uniformly random subpermutation of $\sigma$ of some size $q = q(\varepsilon)$ and makes its decision based on this subpermutation. This can be thought of as sampling a subset $I \subseteq [n]$ of size $q$ and passing $\sigma[I]$ to the tester. We stress that the tester only sees $\sigma[I]$ (i.e., the order pattern of $\sigma|_I$), but not $I$ itself or its image under $\sigma$. 
	Naturally, $q(\varepsilon)$ is called the {\em sample-complexity} of the tester.
	
	This notion of testing is an analogue of the canonical testers (for graph properties) mentioned in Section \ref{sec:introduction}. Unlike in the graph case, where the Goldreich-Trevisan theorem implies that every testable property can be tested by a canonical tester, there are simple permutation properties that cannot be tested by examining subpermutations. For example, as observed in \cite{HKMS}, the property of having at least $\alpha n$ fixed points cannot be tested in this way. On the other hand, it is easy to test this property by just sampling indices and checking if they are a fixed point (clearly, here the tester needs to know the actual ``names" of the elements).
	
	Having said the above, testing via subpermutations is still arguably the most natural way of approaching the testing of {\em hereditary} permutation properties, where a hereditary property is a property closed under taking subpermutations. 
	Note that hereditary properties are exactly those that are characterized by forbidden subpermutations, i.e., for every hereditary property $\mathcal{P}$ there is a family of permutations $\mathcal{F}$ such that a permutation $\sigma$ satisfies $\mathcal{P}$ if and only if $\sigma$ is $\pi$-free for every $\pi \in \mathcal{F}$, where being {\em $\pi$-free} means that $\sigma$ does not contain $\pi$ as a subpermutation. 
	Klimo\v{s}ov\'a and Kr\'al' \cite{KK}, proving a conjecture of Hoppen, Kohayakawa, Moreira and Sampaio \cite{HKMS}, showed that every hereditary permutation property is testable with one-sided error with respect to the Kendall tau distance, i.e., it can be tested with sample-complexity depending only on $\varepsilon$. 
	\begin{theorem}[\cite{KK}]\label{thm:testing Kendall tau}
		Every hereditary permutation property is testable with one-sided error w.r.t.~the Kendall tau distance.
	\end{theorem}
	The proof in \cite{KK} gives Ackermann-type bounds on the sample complexity of testing hereditary properties, even in the case of testing $\pi$-freeness for a single permutation $\pi$. Fox and Wei \cite{FW_conference} later announced a polynomial bound on the sample complexity. 
	\begin{theorem}[\cite{FW_conference}]
		For every permutation $\pi$, $\pi$-freeness is testable with one-sided error w.r.t.~the Kendall tau distance with sample complexity $\poly(1/\varepsilon)$. 
	\end{theorem}

	Another well-known permutation metric is {\em Spearman's footrule distance}, defined as $d_{SF}(\sigma,\pi) = \frac{1}{\binom{n}{2}}\sum_{i=1}^n |\sigma(i) - \pi(i)|$ (where $\sigma,\pi$ are two permutations on $[n]$). It was proved by Diaconis and Graham \cite{DG} that $d_{KT}(\sigma,\pi) \leq d_{SF}(\sigma,\pi) \leq 2d_{KT}(\sigma,\pi)$, and so testing w.r.t.~the Spearman footrule distance is essentially equivalent to testing w.r.t.~the Kendall tau distance.  
	
	We now consider yet another important permutation metric, the {\em rectangular distance}, defined as follows. An {\em interval} is a set of the form $\{x : a \leq x \leq b\}$.
	\begin{definition}[rectangular distance]
		Let $\sigma,\pi$ be permutations on $[n]$. The {\em rectangular distance} $d_{\square}(\sigma,\pi)$ is defined as
		$$
		d_{\square}(\sigma,\pi) = \frac{1}{n} \max_{S,T \subseteq [n]} \big| |\sigma(S) \cap T| - |\pi(S) \cap T| \big|,
		$$
		where the maximum is over all intervals $S,T$ in $[n]$.
	\end{definition}
	The rectangular distance is analogous to the important {\em cut distance} of graphs, which, e.g., underlies the Frieze-Kannan regularity lemma mentioned in Section \ref{sec:estimation} (see the book of Lov\'asz \cite{Lovasz} for \nolinebreak an \nolinebreak overview). 
	
	It can be shown that if the Kendall tau distance between two permutations is small then so is their rectangular distance, but the converse is not true\footnote{For example, the typical rectangular distance between two random permutations is $o(1)$, while their typical Kendall tau distance is roughly $\frac{1}{2}$.}. The fact that small Kendall tau distance implies small rectangular distance means that testing w.r.t.~the rectangular distance is not harder than testing w.r.t.~the Kendall tau distance. Hence, Theorem \ref{thm:testing Kendall tau} implies that every hereditary permutation property is testable w.r.t.~the rectangular distance; this was in fact proven earlier by Hoppen, Kohayakawa, Moreira and Sampaio \cite{HKMS}. 
	
	The rectangular distance can be visualized by viewing a permutation $\sigma$ as the $n$ points $(i,\sigma(i))$ in the square $[n] \times [n]$. For two intervals $S,T$, the quantity $|\sigma(S) \cap T|$ is simply the number of points $(i,\sigma(i))$ within the rectangle $S \times T$. Thus, $d_{\square}(\sigma,\pi)$ being small means that $\sigma,\pi$ have roughly the same number of points (up to $o(n)$) within each such rectangle. 
	
	Fox and Wei \cite{FW_journal} identified another natural permutation metric, called the {\em planar tau distance}, that turns out to be equivalent to the rectangular distance. To define this metric, we first need the following definition. 
	\begin{definition}[planar adjacent transposition]\label{def:planar transposition}
		Let $\sigma,\pi$ be two permutations on $[n]$. We say that $\pi$ is obtained from $\sigma$ by a single planar adjacent transposition if there is $1 \leq i \leq n-1$ such that one of the following holds:
		\begin{itemize}
			\item $\pi(i) = \sigma(i+1)$, $\pi(i+1) = \sigma(i)$, and $\pi(j) = \sigma(j)$ for all $j \neq i,i+1$.
			\item $\pi^{-1}(i) = \sigma^{-1}(i+1)$, $\pi^{-1}(i+1) = \sigma^{-1}(i)$, and $\pi^{-1}(j) = \sigma^{-1}(j)$ for all $j \neq i,i+1$.
		\end{itemize}
	\end{definition}
	The first item in Definition \ref{def:planar transposition} simply says that $\pi$ is obtained from $\sigma$ by an adjacent transposition. When identifying $\sigma$ with the point-set $\{(i,\sigma(i)) : i \in [n]\}$, this corresponds to switching two points that are horizontally adjacent. The second item says that $\pi^{-1}$ is obtained from $\sigma^{-1}$ by an adjacent transposition, which can be thought of as switching two points that are vertically adjacent. 
	
	\begin{definition}[planar tau distance]
		Let $\sigma,\pi$ be two permutations on $[n]$. The {\em planar tau distance} $d_{PT}(\sigma,\pi)$ is defined as $\frac{1}{\binom{n}{2}}$ times the minimum number of planar adjacent transpositions needed to turn $\sigma$ into $\pi$. 
	\end{definition}
	\noindent
	Fox and Wei \cite{FW_journal} proved the following theorem on testing with respect to the planar tau distance.
	\begin{theorem}[\cite{FW_journal}]\label{thm:FW}
		For every hereditary permutation property $\mathcal{P}$ and $\varepsilon > 0$, there is $n_0 = n_0(\varepsilon,\mathcal{P})$ such that the following holds:
		\begin{enumerate}
			\item There exists a constant $C(\mathcal{P})$ such that $\mathcal{P}$ is $\varepsilon$-testable with two-sided error with sample complexity 
			$C(\mathcal{P}) \cdot \tilde{O}(\frac{1}{\varepsilon})$ w.r.t.~the planar tau distance, assuming that input permutations have length at least $n_0$. 
			\item $\mathcal{P}$ is $\varepsilon$-testable with two-sided error with sample complexity $\frac{20000}{\varepsilon^2}$ w.r.t.~the planar tau distance, assuming that input permutations have length at least $n_0$. 
		\end{enumerate}
	\end{theorem}
	
	Note that in the second item, the sample complexity is independent of $\mathcal{P}$. See \cite{FW_journal} for the definition of the constant $C(\mathcal{P})$ from Item 1, as well as for additional results on testing with one-sided error  w.r.t.~the planar tau distance.
	
	Fox and Wei \cite{FW_journal} also showed that $\Omega( d_{\square}(\sigma,\pi)^2 ) \leq d_{PT}(\sigma,\pi) \leq O(d_{\square}(\sigma,\pi)^{1/2})$ for every two permutations $\sigma,\pi$, meaning that the planar tau distance and the rectangular distance are equivalent (up to squaring the distance). Hence, Theorem \ref{thm:FW} also applies to the rectangular distance, but with sample complexity $\tilde{O}(\frac{1}{\varepsilon^2})$ and $O(\frac{1}{\varepsilon^4})$ in Items 1 and 2, respectively. See \cite{FW_journal} for additional permutation metrics that are equivalent to the rectangular distance. 
	
	\paragraph{Acknowledgments.} We thank the anonymous referees for a careful reading of the paper and many useful suggestions that improved the presentation.  
	
	\bibliographystyle{abbrv}
	\bibliography{library}
\end{document}